\documentstyle{natokap}
\begin{opening}
\title{SOLITONS AND LONG JOSEPHSON JUNCTIONS}
\author{R. D. Parmentier}
\institute{Department of Physics, University of Salerno,
I-84081 Baronissi (SA), Italy}
\end{opening}
\begin{document}
\section*{}
\begin{center}
\centering{\dag This paper is dedicated to\\the memory of Gianfranco
Patern\`{o}.}
\end{center}
\vspace{0.8cm}

\begin{abstract}
Magnetic flux quanta, of value $h/2e$, in long Josephson junctions behave as
(quasi) solitons. Fluxon dynamical states are well described by a perturbed
sine-Gordon equation model, with boundary conditions determined by the
junction geometry and by externally applied magnetic fields, and they give
rise to readily measurable physical phenomena, such as step structure in
current-voltage characteristics and microwave radiation emission. Devices
based on fluxon propagation offer potentially interesting applications as
oscillators and amplifiers---as well as digital applications, described
elsewhere in this volume---in high performance integrated superconductive
circuits.
\end{abstract}
\section{Introduction}
The combination of `solitons' and `long Josephson junctions' brings
together the interests of a broad and varied mix of mathematicians,
physicists and device engineers. Although the underlying mathematical ideas
had been around for some time, the term `soliton' itself was coined in 1965
by Zabusky and Kruskal [1], just three years after the original work by
Josephson [2]. It required some time, however, for the practical importance
of the soliton concept to be fully appreciated by the applied science
community---certainly more time than was required to absorb the
significance of Josephson's work. During the late 1960's and early 1970's
`solitons' and `long Josephson junctions' began to attract increasing
research attention. Starting from key papers by Kulik [3], Scott [4], and
Fulton and Dynes [5] (I warn the reader that my literature citations are
subjective, partial and incomplete), the literature gradually developed,
bringing about an improved level of understanding of the basic dynamics of
solitons in long junctions. A landmark paper that was particularly
influential in focusing interest on the significance of solitons in
physical systems was Scott {\em et al.} [6]. About the same time, a number
of people (see, {\em e.g.}, Fulton {\em et al.} [7]) began to realize that
solitons in long junctions offer an exciting potential for practical device
applications. In the intervening years research in the field has continued
unabated, both `basic' research, with the development of refined
approximate analytical tools and an ever more detailed exploration of
`exotic' topics such as chaos, and `applied' research, with the development
of novel devices such as phase locked arrays of oscillators and the vortex
flow transistor.

In this present review my essential task is to attempt an up-date of
material presented in two previous reviews, one written by me in 1978 [8]
and the other by Niels Pedersen in 1986 [9]. To make this present chapter
reasonably coherent and self-consistent it will be necessary to repeat a
certain amount of material already presented in these two previous reviews;
my main emphasis, however, will be on material that has appeared later.
\section{Device Geometries and Models}
\subsection{Geometries}
By `long Josephson junction' we mean simply one that has one dimension,
along, say, the $x$-coordinate, long with respect to the Josephson
penetration length, $\lambda_{J}$ (see, {\em e.g.}, Bruynseraede {\em et
al.} [10] for a concise description of the fundamentals of Josephson
tunneling; for a more detailed description see Barone and Patern\`{o} [11]),
and the other dimension, along, say, the $y$-coordinate, short with respect
to $\lambda_{J}$. One might well ask why we choose to focus attention on
what is, after all, a rather particular geometrical configuration. I
suspect that the true answer here is more a question of analytical
convenience than it is of physical importance: a `small' Josephson junction
is described mathematically by an ordinary differential equation (ode); a
`long' junction is described by a partial differential equation (pde) in
one space dimension and time; a `large' (in both the $x$- and $y$-
dimensions) junction is described by a pde in two space dimensions and
time. A $(2+1)$-dimensional pde is a much more complicated object than is a
$(1+1)$-dimensional pde.

Long Josephson junctions can be constructed in different geometries, and
the junction geometry has a significant effect on the junction dynamics.
The most extensively studied geometries are the overlap geometry, the
inline geometry and the annular geometry [9]. The first two are somewhat
similar in that they both consist of finite-length, quasi-one-dimensional
strips; the essential difference between them is the manner in which the
bias current is applied to them: in the overlap geometry the current is
applied perpendicular to the long dimension of the junction, whereas in the
inline geometry it is applied parallel to the long dimension. In fact,
overlap and inline are idealized limiting cases; real physical
finite-length junctions are almost always some combination of the overlap
and the inline geometries [9]. The annular geometry, instead, is
qualitatively different, consisting of a strip closed upon itself in an
annulus. Although somewhat more difficult to construct and control from the
point of view of fabrication technology, the annular geometry junction
offers the interesting possibility of studying soliton dynamics in the
absence of boundary reflection effects.
\subsection{Mathematical Models}
The electrodynamics of a Josephson junction is, in general, described by a
$(2+1)$-dimensional pde for the phase difference $\phi$ between the
junction electrodes. In normalized form this equation can be written [12]
\begin{equation}
\phi_{xx}+\phi_{yy}-\phi_{tt}-\sin\phi = \alpha\phi_{t}-\beta
(\phi_{xxt}+\phi_{yyt})
\end{equation}
(we ignore here the so-called $\cos\phi$ term). Here, $x$ and $y$ are the
spatial coordinates normalized to $\lambda_{J}$, $t$ is time normalized to
the inverse of the Josephson angular plasma frequency $\omega_{J}$
($\lambda_{J} \omega_{J} = \overline{c}$, the maximum electromagnetic
propagation velocity in the junction, called also Swihart velocity), the
term in $\alpha$ represents shunt dissipation due to quasi-particle
tunneling (normally assumed ohmic, but see Scheuermann and Chi [13] for the
effect of a physically more realistic, non-ohmic term), the term in $\beta$
represents dissipation due to the surface resistance of the superconducting
junction electrodes [14], and subscripts denote partial derivatives.
Details of the normalizations may be found in Ref. [12]. Boundary
conditions for Eq. (1), at $x = 0,L$ and $y = 0,W$, are determined,
respectively, by the $y$- and $x$-components of the magnetic field to which
the junction is subjected [11].

For the case of a `long' junction, {\em i.e.}, one for which $L \gg 1$ and
$W \ll 1$, it is relatively simple to calculate the relevant component of
the magnetic field generated by the junction bias current [11,12]. For the
overlap geometry, assuming a constant (physical) bias current $I_{bias}$ in
the $y$-direction and a spatially uniform, but possibly time-varying,
external magnetic field applied also in the $y$-direction, the model
reduces to
\begin{equation}
\phi_{xx}-\phi_{tt}-\sin\phi = \alpha\phi_{t}-\beta\phi_{xxt}-\gamma\,,
\end{equation}
with the boundary conditions
\begin{equation}
\phi_{x}(0,t)+\beta\phi_{xt}(0,t)=\phi_{x}(L,t)+\beta\phi_{xt}(L,t)=\eta\,.
\end{equation}
Here, $\gamma \equiv I_{bias}/J_{o} L \lambda_{J}$, where $J_{o}$ is the
(physical) maximum pair-current density per unit length in the
$x$-direction, and $\eta$ is a normalized measure of the $y$-component of
the external magnetic field (the normalization is such that, with $\gamma =
0$, $\eta = 2$ is the static threshold for magnetic flux entry into the
junction). A condition for the validity of the reduction to one dimension
is [12]: $\gamma W^{2}/8 \ll 1$. For the inline geometry the bias enters
through the boundary conditions rather than as a term in the pde, since
$I_{bias}$ is now applied in the $x$-direction. Thus, the pde becomes
\begin{equation}
\phi_{xx}-\phi_{tt}-\sin\phi = \alpha\phi_{t}-\beta\phi_{xxt}\,,
\end{equation}
and the boundary conditions become
\begin{eqnarray}
\phi_{x}(0,t)+\beta\phi_{xt}(0,t) = -\kappa + \eta\,,\\
\phi_{x}(L,t)+\beta\phi_{xt}(L,t) = +\kappa + \eta\,,
\end{eqnarray}
where $\kappa \equiv I_{bias}/2 J_{o} \lambda_{J}$. Finally, for an annular
junction in the absence of an externally applied magnetic field, assuming a
spatially uniform bias current density, the governing pde is once again Eq.
(2), where $x$ is now to be interpreted as the dimension along the mean
circumference of the annulus, and the boundary conditions simply reflect
the spatial periodicity of the system:
\begin{equation}
\phi (x+L,t) = \phi (x,t) + 2 \pi n \, ,
\end{equation}
where $n$ is an integer (the physical significance of which is the
difference between the number of solitons and the number of antisolitons
trapped in the junction; it is also known as the `winding number', for
reasons that will become clear during the discussion of the mechanical
analog). In the presence of an externally applied magnetic field the right
hand side of Eq. (2) must be augmented by the term $\Delta
\frac{\partial}{\partial x}(\vec{B} \cdot \vec{r})$, as discussed by
Gr\o nbech-Jensen [15], where $\vec{B}$ is the normalized applied magnetic
induction, $\vec{r}$ is a unit normal (radial) vector, and $\Delta$ is a
coupling constant.
\subsection{Mechanical Models}
The analogy between the dynamical equation describing a small Josephson
junction and that describing a simple plane pendulum subjected to viscous
damping and an applied mechanical torque is well known and has been
utilized by many authors. The extension of this analogy to the long
junction case was quickly recognized and exploited---see, {\em e.g.}, Scott
[4], Fulton [16], and Cirillo {\em et al.} [17]; the extension consists of
an elastic coupling between adjacent pendula, thus contributing a
`$\phi_{xx}$' term to the dynamical equation; a very simple `pocket
calculator' version can easily be constructed by inserting dress-maker's
pins into a rubber band at regular intervals.

In spite of the development of high-speed digital computers in recent
years, it is my opinion that the mechanical model of the long junction
remains still today an extremely effective stimulus for physical intuition.
It renders immediately evident, for example, the nature of the three
fundamental classes of excitations on long junctions, {\em viz.}, plasma
oscillations, breathers and kinks---a fact that was recognized, and
carefully illustrated, by Fulton [16]. Plasma oscillations, called also
`phonons' or `radiation' in other contexts, stem from oscillatory motions
of the single pendulum (see Fig. 20 of [16]). The lowest-order, $k = 0$
($k$ is the wave number), mode has all of the pendula in the chain
oscillating in a spatially uniform motion; higher-order, $k > 0$, modes have
a standing wave structure, with one or more spatial nodes along the
pendulum chain. For the pure sine-Gordon equation (Eq. (2) with loss and
bias terms set to zero), small-amplitude plasma oscillations are
characterized by the dispersion relation ${\omega}^{2} = k^{2} + 1$; thus,
$\omega = 1$ is the lowest possible frequency for such oscillations.
Breather oscillations, called also `bions' in earlier literature, also stem
from oscillatory motions of the single pendulum (see Fig. 28 of [16]). They
differ from plasma oscillations in that the oscillation amplitude varies
with position along the chain, but there are no spatial nodes along the
chain. For the pure sine-Gordon system of infinite length, breathers can
have frequencies in the range $0 < \omega < 1$. In contrast, kink
oscillations stem from rotary, `over-the-top', motions of the single
pendulum (see Figs. 25 and 26 of [16]); a single isolated kink is just a
$2\pi$ twist in the pendulum chain. Viewing Fulton's careful illustrations
also renders intuitive the idea that a breather may be considered to be an
oscillatory bound state of a kink and an anti-kink, an idea that has
significant consequences for sine-Gordon dynamics. In fact, thoughtful
study of Fulton's figures was instrumental in deriving the corresponding
exact analytic solutions of the pure sine-Gordon equation on the finite
interval, for arbitrary oscillation amplitudes, by Costabile {\em et al.}
[18].

At this point I should mention that whereas both breathers and kinks of
the pure sine-Gordon equation are properly classified as `solitons' from a
strictly mathematical point of view, people who study applications of long
Josephson junctions normally reserve the term `soliton' to mean only kinks.
The reason for this is undoubtedly that, due to the dissipative
terms---those in $\alpha$ and $\beta$---in the long junction model, plasma
oscillations and breathers can exist only as short-lived transients, unless
very special forms of driving terms are employed, whereas kinks are highly
robust objects that can emerge from a power balance between the $\alpha$-
and $\beta$-dissipative terms and even a simple constant bias current.
Kinks are also often called `fluxons' in the Josephson junction context
because each one contains exactly one quantum of magnetic flux, $h/2e$ ($h$
is Planck's constant and $e$ the electron charge); they are also called
`vortices' because the electric current surrounding a fluxon has the form
of a vortex. I shall, in the following, use interchangeably---even if
somewhat improperly---the terms `soliton', `kink', `fluxon', and `vortex'
to refer to these robust, particle-like objects. Finally, the significance
of the `winding number' $n$ in Eq. (7) is now clear: if we introduce $n$
complete $2 \pi$ twists, {\em i.e.}, `kinks', into the pendulum chain before
closing the two ends into an annulus, the number $n$ is evidently a
topological invariant of the system.
\section{Soliton Dynamics}
\subsection{Basic Phenomena}
Let us suppose now that we have introduced---never mind how (for the
moment)---a single kink into the interior of a long junction. How will this
object behave? The answer evidently depends on the type of junction
involved, {\em i.e.}, the boundary conditions, and on the values of the
various parameters of the system; the mechanical model will serve as an aid
to the imagination here.

To be specific, suppose that we have an overlap geometry junction with
$\alpha \ll 1$ and $\beta \ll 1$, in the absence of an external magnetic
field, {\em i.e.}, $\eta~=~0$, with a bias current, $0 < \gamma < 1$. In
the absence of the kink, the bias current, {\em i.e.}, torque, causes a
uniform angular offset of the static equilibrium position of the pendulum
chain, of value $\phi = \arcsin\gamma$. When the kink is introduced, this
applied torque pushes it toward one end of the junction, which end
depending on the sense of winding of the kink with respect to that of the
applied torque. With $\eta = 0$ the two ends of the pendulum chain are
free. If the kink arrives at the junction end with enough kinetic energy,
the pendulum chain will continue to wind up, {\em i.e.}, the end pendulum
will undergo a rotation of $2\pi + 2\pi = 4\pi$, and the resulting
anti-kink will be pushed toward the other junction end, where the
reflection mechanism will repeat. If, on the other hand, the kink's kinetic
energy is not sufficient to survive the reflection, {\em i.e.}, the end
pendulum does not receive enough energy to continue winding up, the kink
will `die' at the first end, and, after a transient oscillation, the
junction will relax to the spatially uniform static state, $\phi =
\arcsin\gamma$. The former case evidently leads to a dynamic steady state,
with the kink bouncing resonantly back and forth across the junction, and
each pendulum in the chain increasing its angular position by $4\pi$ after
each complete back and forth period of the kink. Since angular velocity,
$\phi_{t}$, in the mechanical model corresponds to voltage in the junction,
a state with $<\phi_{t}> \neq 0$ should be manifested as structure in the
current-voltage characteristic of the junction, a fact recognized by
Fulton and Dynes [5], who named such structures `zero-field steps'. Since a
kink behaves in much the same way as does a relativistic particle having a
(normalized) limiting velocity equal to unity (physical limiting velocity
$= \overline{c}$), it follows that these steps tend asymptotically to a
constant voltage. An experimental current-voltage characteristic of an
overlap geometry junction, showing six such zero-field steps, is shown in
Fig. 1. The step index corresponds to the number of fluxons participating
in the dynamic state; the internal symmetry of a multi-fluxon state can
obviously be more or less complicated, but the basic dynamic mechanism
remains that of resonant back and forth propagation.
\begin{figure}
\vspace{7.0cm}
\caption[]{Experimental current-voltage characteristic of a long Josephson
junction of overlap geometry, showing six zero-field steps. The step index
corresponds to the number of fluxons participating in the dynamics.}
\end{figure}

Let us consider now the question, left unanswered above, of how to
introduce a fluxon into an experimental junction. To observe zero-field
steps, the experimentalist increases the bias current from zero up to the
maximum critical value, at which point the junction switches from the
zero-voltage state to the gap-voltage state. The equivalent pendulum chain
switches from the static state, $\phi = \arcsin\gamma$, to a spatially
uniform rotating state when $\phi$ reaches $\pi /2$. The experimentalist
then progressively reduces the bias current, tracing out the McCumber
background curve of the characteristic, a piece of which is indicated as
`MCB' in Fig. 1. Correspondingly, the pendulum chain rotates at an ever
slower angular velocity. If the bias current is then increased again before
the junction returns to the zero-voltage state, it is possible to trace out
the zero-field steps, as shown in Fig. 1. The basic mechanism involved was
noted by Fulton [16] through observation of the dynamics of a pendulum
chain and elaborated analytically only some time later by Costabile {\em et
al.} [19]: at a sufficiently slow rotation velocity the pendulum chain can
develop spontaneously a dynamic instability that results in the `birth' of
one or more kinks.

Suppose now that to the situation described above we add a small dc
magnetic field, {\em i.e.}, $0 < \eta \ll 1$. This term corresponds to a
small torque applied to just the two end pendula of the chain; at one end
the $\eta$-torque is in the same sense as the $\gamma$-torque, and at the
other end the two torques are opposed. The consequence of adding this term
is immediately apparent: at the end where the two torques add, the
reflection of fluxons is enhanced, {\em i.e.}, energy is added to the
system; at the other end energy is subtracted from the system, and fluxon
reflection is impeded. Increasing gradually the value of $\eta$, this
asymmetry between the two junction ends is progressively accentuated until,
with an appropriate combination of parameters, a different dynamic steady
state configuration can ensue: when the fluxon reaches the
energy-subtracting end it `dies', giving rise to a packet of
small-amplitude oscillations (a more or less complicated combination of
breathers and plasma waves), which in turn propagates back toward the
energy-adding end; at that end, the energy of the packet is sufficient to
`give birth' to a new fluxon, which again propagates toward the
energy-substracting end, {\em etc}. This is the mechanism responsible for
the appearance of `Fiske steps' in long Josephson junctions [20]. In the
dynamic state corresponding to the first zero-field step each pendulum in
the chain advances by $4\pi$ during one complete period; in the dynamic
state corresponding to the first Fiske step each pendulum advances by
$2\pi$ during one complete period. This implies that the asymptotic
limiting voltage of the first Fiske step is just one-half that of the first
zero-field step. Analogously to the situation for zero-field steps, there
also exist higher-order Fiske steps having a progressively more complicated
internal structure [20].

As the value of $\eta$ is increased still further, yet another dynamic
steady state configuration emerges: that of flux flow. As mentioned above,
$\eta = 2$ is the static threshold for the penetration of magnetic flux
into the junction in the absence of a bias current. With $\gamma > 0$, this
threshold value is reduced (at the energy-adding end); furthermore, a
fluxon injected into the junction at this end is accelerated into the
interior of the junction by the bias current. When the combination of
$\eta$ and $\gamma$ is appropriately above threshold, a dynamic steady
state configuration ensues in which fluxons are injected at the
energy-adding end as fast as the junction can `swallow' them; these are
accelerated toward the other junction end, where they may be reflected or
annihilated, depending on the boundary conditions at that end. A computer
simulation illustrating this situation is shown in Fig. 2. The average
junction voltage in such a state is proportional to the average number of
fluxons inside the junction times their average velocity. Once again, since
a fluxon behaves much like a relativistic particle, the current-voltage
characteristic tends toward a constant voltage as the average fluxon
propagation velocity approaches its limiting value. The resulting steps in
the characteristic are commonly called `flux-flow steps' or
`velocity-matching steps'. The potential interest of flux flow dynamics for
practical electronic device applications was also recognized quite
early---see, {\em e.g.}, Yoshida and Irie [22].
\begin{figure}
\vspace{6.0cm}
\caption[]{Computer simulation showing instantaneous junction voltage in
the flux flow regime for parameter values: $L = 30, \alpha = 0.25, \beta =
0.001, \gamma = 1.28, \eta = 4$. After Zhang [21] (reprinted with
permission).}
\end{figure}

The basic dynamical phenomena in an inline geometry junction are
qualitatively quite similar to those in an overlap geometry junction, {\em
viz.}, zero-field steps, Fiske steps, and flux flow. Quantitative
differences are due to the difference in the effects of the bias current.
Whereas in an overlap junction each pendulum in the chain is subjected to a
torque corresponding to $\gamma$, in an inline junction only the two end
pendula are subjected to a torque corresponding to $\kappa$, according to
Eqs. (5) and (6). In the absence of kinks, each pendulum in an overlap
chain is offset in static equilibrium position by the same angle, $\phi =
\arcsin\gamma$, whereas the angular offset relaxes toward zero over a
physical distance $\sim\lambda_{J}$ (normalized distance $\sim 1$)
proceeding from the ends toward the interior of an inline chain; the
corresponding analytic solution for this latter case was derived by Owen
and Scalapino [23] in terms of Jacobian elliptic functions. The most
significant consequence of this difference for fluxon dynamics is that a
fluxon in an overlap junction feels a constant accelerating force due to
$\gamma$ along the entire length of the junction, whereas a fluxon in an
inline junction receives an accelerating `kick' due to $\kappa$ only near
the two junction ends.

The situation for an annular geometry junction is quite different because
there are no junction ends and because the winding number is fixed for a
given dynamical configuration; this means, {\em e.g.}, that Fiske steps are
not present, and that the distinction between zero-field steps and
velocity-matching steps tends to be obscured. On the other hand, the
possibility of studying fluxon dynamics in the absence of boundary
reflection effects is often of considerable interest; in this sense the
annular geometry offers a `cleaner' environment than does either the
overlap or the inline geometries. An interesting feature of fluxon dynamics
in annular junctions is that for winding numbers different from zero and in
the absence of external magnetic fields there is no zero-voltage current in
the current-voltage characteristic of the junction. The reason for this is
that as soon as a $\gamma > 0$ is applied, any fluxon that is locked into
the junction begins to move, which implies that $<\phi_{t}> \neq 0$. This
situation changes if a small dc magnetic field is applied in the plane of
the junction: As mentioned in Sect. 2.2, the presence of an externally
applied magnetic field is modelled by adding the term $\Delta
\frac{\partial}{\partial x}(\vec{B} \cdot \vec{r})$ to the right hand side
of Eq. 2. If $\vec{B}$ is time-independent and spatially homogeneous, this
term gives rise to a spatially sinusoidal potential function which can
effectively pin fluxon motion [24,15]. Very recently, Ustinov {\em et al.}
[25] have devised an ingenious technique for adding fluxons in a controlled
way into an annular junction; this will undoubtedly stimulate further
experimental study of this geometry.
\subsection{Soliton Perturbation Theory}
The only completely general technique (aside from experimental measurement)
for studying soliton dynamics in long junctions is that of detailed
simulation, {\em i.e.}, numerical integration of the appropriate pde model.
It is frequently convenient, however, to be able to employ approximate
techniques to obtain approximate, even though incomplete, information.
Soliton perturbation theory furnishes such a technique. Since this topic
was discussed quite thoroughly in the 1986 review by Pedersen [9] I shall
limit myself here simply to recalling a few salient points. The interested
reader should consult McLaughlin and Scott [26] for the mathematical basis
of this theory, and the review by Kivshar and Malomed [27] for a detailed
discussion of applications to long Josephson junctions.

The basic physical idea involved is that a kink is a highly robust,
particle-like object that, even though acted upon by perturbations, tends
to maintain its identity; the mathematical foundation underlying this
physical fact is that the pure sine-Gordon equation is an exactly
integrable system [6]. Accordingly, it is possible to describe the dynamics
of this object in terms of its energy, defined as
\begin{equation}
H = \int[\frac{1}{2}\phi_{x}^{2}+\frac{1}{2}\phi_{t}^{2}+(1-\cos\phi)]\,dx
\end{equation}
and its momentum, defined as
\begin{equation}
P = -\int\!\phi_{x}\phi_{t}\,dx\,.
\end{equation}
The approximation involved in using the soliton perturbation approach
derives from the fact that, since we obviously do not know the unknown
solution of the perturbed equation, we use instead the known solution of
some `nearby' unperturbed equation in Eqs. (8) and (9). As a simple
illustration of this procedure, let us consider the problem of describing
the motion of a single fluxon in an overlap junction, described by Eq. (2),
of infinite length. In this case, the time derivative of Eq. (8) may be
calculated as
\begin{equation}
\frac{dH}{dt} = \int_{-\infty}^{+\infty}[\gamma\phi_{t}-\alpha\phi_{t}^{2}
-\beta\phi_{xt}^{2}]\,dx\,.
\end{equation}
We use as a solution ansatz the function
\begin{equation}
\phi(x,t) = 4\,\arctan\,[\exp(-[x-X(t)]/[1-u^{2}(t)]^{1/2})]\,,
\end{equation}
which, with $X(t)=ut$ and $u=constant$, is just the well-known single
soliton solution of the pure sine-Gordon equation on the infinite interval.
For this unperturbed problem, the energy of the solution described by Eq.
(11) may be calculated from Eq. (8) as
\begin{equation}
H = 8/(1-u^{2})^{1/2}\,.
\end{equation}
By using the ansatz of Eq.(11), we are implicitly assuming that the
essential effect of the perturbing terms---those in $\alpha$, $\beta$, and
$\gamma$---on the dynamics of the system is to cause a slow modulation of
the parameters $X$ and $u$. This modulation is obtained by inserting Eq.
(11) into Eq. (10), and using Eq. (12), which gives the result
\begin{equation}
\frac{du}{dt} = \frac{\pi\gamma}{4}\,(1-u^{2})^{3/2}-\alpha u\,(1-u^{2})
-\beta u/3
\end{equation}
and
\begin{equation}
X(t) = x_{o} + \int_{t_{o}}^{t}\!u(\tau)\,d\tau\,,
\end{equation}
where $x_{o}$ is the position of the fluxon at time $t_{o}$. In this way,
we have reduced the problem from the integration of a pde to the
integration of an ode, which provides a substantial computational
simplification.

The accuracy of results obtained using the perturbation theory approach
depends on how `close' (in some function-space sense) the solution of the
perturbed equation is to the solution of the unperturbed equation used to
approximate it in the energy and momentum integrals. A practical example in
which the perturbation approach breaks down due to significant
discrepancies between the two solutions, caused by the effects of the
$\beta$-loss term, was described in detail by Davidson {\em et al.}
[28]. The only generally valid way to verify the accuracy of perturbation
theory results is to compare them {\em a posteriori} with pde simulation
results. In spite of this inherent weakness, the perturbation theory
approach has repeatedly proved to be an extremely useful tool for the study
of soliton dynamics.
\subsection{Time-Dependent Drivers}
So far, we have considered only constant, time-independent bias currents
and magnetic fields. In many cases of practical interest, however, the
behavior of a long junction in the presence of a time-varying driver is of
considerable interest. As a simple illustration of this class of problems,
let us consider the inline junction described by Eqs. (4)-(6), subjected to
a sinusoidally varying magnetic field, {\em i.e.}, with $\eta$ in Eqs. (5)
and (6) given by $\eta = \eta_{o}\sin (\omega t + \theta)$; this might
model a situation in which such a junction is exposed to a microwave field.

The perturbation theory developed in the previous section offers a
convenient tool for analyzing this problem. For simplicity we will treat
only the case $\beta = 0$, which was solved by Salerno {\em et al.} [29];
the general case, with $\beta \neq 0$, was solved by Filatrella {\em et
al.} [30].

For an inline junction with $\beta = 0$ the time derivative of Eq. (9) is
readily calculated to be
\begin{equation}
\frac{dP}{dt} = -\alpha P\,,
\end{equation}
the general solution of which is obtained by a trivial integration.
Moreover, for the unperturbed equation, we have that $P = uH$, with $H$
given by Eq. (12). From these facts the soliton trajectory may be found
explicitly from Eq. (14) as
\begin{equation}
X(t) = x_{o} -
\frac{1}{\alpha}\ln
\left[\frac{z+(z^{2}+1)^{1/2}}{z_{o}+(z_{o}^{2}+1)^{1/2}}\right]\,,
\end{equation}
where $z \equiv P/8$ and $z_{o} \equiv z(t_{o})$.

The other essential ingredient of the analysis is the treatment of fluxon
reflections at the boundaries. This problem was solved in the context of
the perturbation theory by Levring {\em et al.} [31], who showed that
during a boundary reflection, due to Eqs. (5) and (6) (with $\beta = 0$), a
fluxon undergoes an energy variation $\Delta E$, given by
\begin{equation}
\Delta E = 4 \pi (\kappa \pm \eta)\,,
\end{equation}
where the plus sign is taken at one boundary and the minus sign at the
other. This boundary effect can be incorporated into the description of the
system dynamics very simply by rewriting Eq. (15) as
\begin{equation}
\frac{dP}{dt} = - \alpha P + 4 \pi \sum_{k=0}^{\infty}[\kappa + (-1)^{k+1}
 \eta_{o} \sin (\omega t + \theta)]\,\delta (X(t)-kL)\,,
\end{equation}
in which $\delta ()$ is the Dirac delta and $X(t)$ is given by Eq. (16). In
writing Eq. (18) we are constructing a mathematical artifice consisting of
a periodically extended junction structure lying along the positive
$x$-axis, between $0$ and $+\infty$, in such a way that the back and forth
shuttling motion of the fluxon in the physical junction is transformed into
a unidirectional, left-to-right motion on the extended structure, with
boundary reflection effects taking place at spatial points equal to integer
multiples of $L$.

As shown in detail by Salerno {\em et al.} [29], the system description can
be simplified still further, from the ode of Eq. (18) to an explicit,
two-dimensional functional map whose variables are $t_{k}$, the time,
modulo the period of the applied microwave field, of the $k$'th boundary
reflection, and $E_{k}$, the fluxon energy at the $k$'th reflection. In
this way we reduce enormously the computational difficulty of the problem:
from integrating the pde system of Eqs. (4)-(6), to integrating the ode of
Eq. (18), to iterating a two-dimensional functional map.
\begin{figure}
\vspace{8.2cm}
\caption[]{Current-voltage characteristic of inline junction with
ac magnetic field. Smooth curve: no field; discontinuous curve: with field.
Fundamental frequency: $n = m = 1$. Parameters: $\alpha = 0.05, L = 12,
\eta_{o} = 0.4, \omega = 0.225$. Arrow represents switching to the
zero-voltage state due to fluxon annhilation.}
\end{figure}

The new phenomenon that emerges in this case is that of phase locking: over
a certain range of parameters the back and forth motion of the fluxon in
the junction becomes locked in phase to the external microwave field; this
is most readily manifested as the appearance of a constant-voltage current
step in the current-voltage characteristic of the junction, at a voltage $V
= \frac{2n}{m} \omega$, where $\omega$ is the angular frequency of the
applied field, and $n$ and $m$ are integers, as shown in Fig. 3. Reduction
of the system dynamics to a functional map is particularly convenient for
studying the phase locking phenomenon inasmuch as one can use standard,
well known techniques to study the existence and stability of fixed points
of the map, which corresponds directly to studying the existence and
stability of phase-locked states of the fluxon dynamics. The salient
features that emerge from such an analysis, applied to the system described
by Eq. (18), are [29]: (a) Phase-locked steps similar to the one shown in
Fig. 3 exist at the fundamental frequency $(n = m = 1)$ and at all odd
subharmonics $(n = 1; m = 1, 3, 5, \ldots )$. (b) The height in current of
such a step is $2 \eta_{o}$, {\em i.e.}, equal to the peak-to-peak
amplitude of the ac magnetic field; the step is centered around the
unperturbed current-voltage characteristic. (c) The stability range in
field amplitude for the steps decreases rapidly with increasing subharmonic
order, {\em i.e.}, with increasing $m$. (d) Instability appears first at
the center of a step and not, as one might expect, at the extremities. (e)
For field amplitude values beyond the stability limit, a Feigenbaum-like
bifurcation cascade leading to chaos in the fluxon times of flight across
the junction is observed. (f) For appropriate model parameters the phase
locking steps can cross the zero-current axis, extending to negative
current values; such zero-crossing steps are normally associated with a
hysteresis in the step amplitude.

To check the validity of the particle-map formalism used here, it is useful
to compare results with those obtained from a full simulation of the
original pde model, Eqs. (4)-(6), which has a fairly long and well tested
history of accounting for experimental results. Such a comparison shows
that the strongest limitation of the particle-map approach stems from the
assumption of single fluxon dynamics. Physically, a sufficiently large
energy exchange term, Eq. (17), can give rise to the creation of additional
fluxons in the junction. Such effects are clearly manifested in pde
simulations [32], but they are not contained in the particle-map model;
their principal consequence is that the linear growth of the step height in
current with $\eta_{o}$ holds only up to a certain limit, beyond which the
step height saturates, or even decreases, with increasing field amplitude.
Quantitative discrepancies between map and pde results can often be
attributed to the fact that a fluxon in the particle-map formalism is a
point particle with no spatial extension, whereas a pde fluxon has a
non-zero width. Consequently, whereas energy exchange at the junction
boundaries occurs instantaneously in the map, the exchange interaction is
smeared out over a certain time interval in the pde. This difference tends
to disappear for junction lengths $L \gg 1$, but it may be significant for
length values of practical interest, especially when subharmonic locking is
considered; this problem has been considered in detail, in the context of a
perturbation-theory approach, by Gr\o nbech-Jensen [33]. Nonetheless,
within its range of validity, the simple map approach predicts results to a
truly surprising level of detail and accuracy.
\subsection{Chaos}
As mentioned briefly in the previous section, solitons in long junctions
can also undergo chaotic motion. This is a topic that has attracted a
large amount of attention in recent years, and I refer the reader to the
literature (see, {\em e.g.}, Guerrero and Octavio [34], plus references
therein) for detailed discussions; I present here only a few particular
points. Chaotic effects are observed especially when a junction is
subjected to time-varying currents and/or magnetic fields, but the presence
of such an ac driver is not strictly necessary to bring about chaos:
Soerensen {\em et al.} [35] described computer simulations in which a
junction that is dc-biased in the Fiske-step regime could, with appropriate
parameter values, exhibit chaotically intermittent switching between the
dynamic states corresponding to the first and the second Fiske steps;
further confirmation and analysis of this phenomenon were subsequently
reported by Yeh {\em et al.} [36] and by Filatrella {\em et al.} [37].
\begin{figure}
\vspace{8.2cm}
\caption[]{Map calculation showing dependence of fluxon time of flight on
field amplitude for inline junction with ac magnetic field at third
subharmonic. Parameters: $\alpha = 0.05, L = 10, \kappa = 0.1295, \omega =
0.3$.}
\end{figure}

A case in which chaos is brought about by the presence of an ac driver is
shown in Fig. 4. This figure shows how the fluxon time of flight, defined
as $T_{k} \equiv t_{k} - t_{k-1}$, changes with the amplitude of the
applied magnetic field, for an inline junction similar to that discussed in
the previous section, using the particle-map formalism [29]; that the
phenomenon is real and not just an artifact of the map formalism was
demonstrated analytically by Malomed [38] and and via pde simulation by
Rotoli and Filatrella [39]. The field value at which the first bifurcation
in Fig. 4 occurs is just the one for which a stability analysis indicates
that the phase-locked state loses stability [29]; the values for the
subsequent bifurcations follow quite closely those of a Feigenbaum cascade
[40]. An interesting fact that emerges from the map calculations is that
throughout the bifurcation region, and even for some distance into the
chaotic region, the average junction voltage remains exactly that of the
phase-locked state, below the bifurcation tree, even though the fluxon is
certainly no longer locked to the driver; this fact may be established by
calculating (numerically) the average time of flight in these regions. This
means that, in these regions, the complication in the underlying dynamics
leaves no signature in the current-voltage characteristic of the junction.
Consequently, the only way to reveal such phenomena experimentally would be
through a spectral analysis of the radiation emitted by a junction in a
state involving fluxon propagation; this poses an interesting challenge to
experimentalists.

For most practical electronic device applications the presence of chaotic
phenomena is highly undesirable inasmuch as these constitute simply another
source of electrical noise, thereby degrading some desired performance
characteristic. For this reason, researchers are pressed not only to
identify parameter regions where chaos might exist, in order to avoid them,
but in fact to devise stratagems to suppress chaos. A step in this
direction was recently reported by Salerno [41], who showed, using the
particle-map formalism, that the type of chaos described in the preceding
paragraph can indeed be suppressed by the addition of a small subharmonic
component to the ac magnetic field driving term. Filatrella {\em et al.}
[42] have provided further credibility for this result via pde simulation;
in particular, they showed that, in a situation very similar to that
indicated in Fig. 4, the addition of an $m = 3$ subharmonic component, {\em
i.e.}, a component whose frequency would correspond to the fundamental for
a step at that voltage, approximately one order of magnitude smaller in
amplitude than the (true) fundamental, can be sufficient to suppress chaos
of the type indicated in Fig. 4.
\section{Electronic Applications}
\subsection{Fluxon Oscillators}
Soliton propagation phenomena in long Josephson junctions, both resonant,
back and forth propagation and unidirectional flux flow, give rise
naturally to practical applications as electronic oscillators. The reason
for this is that a soliton is---amonst other things---an electromagnetic
pulse, which, when it impinges on the free end of a junction, gives rise to
the emission of a pulse of electromagnetic energy. If the arrival of
solitons at the junction end is time-periodic, then so is the radiation
emission. For typical experimental junctions made with currently available
fabrication technologies, {\em i.e.}, based on low critical temperature
materials, this radiation lies somewhere within the microwave to
millimeter-wave region of the electromagnetic spectrum. The power emitted
is, in absolute terms, rather low: in this context, one microwatt is a
`large' power; on the other hand, existing devices such as SIS mixers, that
might conceivably be employed in integrated, all-Josephson millimeter-wave
receivers, require not more than this amount of power.

Comparison of the relative merits of oscillators based on the two
propagation modes, {\em viz.}, resonant back and forth propagation and
unidirectional flux flow, shows that they have complementary strengths and
weaknesses. Single-junction resonant propagation oscillators typically show
very narrow output linewidths---about $1$ kHz at oscillation frequencies of
$\sim 10$ GHz have been measured [43]---but observed output powers into an
external load have ranged from picowatts up to something less than about a
nanowatt---really too small to be practically useful (however, Cirillo {\em
et al.} [44] have estimated a power of $\sim 100$ nW at $75$ GHz into a
tightly coupled small junction used as a detector).  In contrast, for
single-junction flux flow oscillators, powers of about a microwatt (into a
small-junction detector) at frequencies up to $\sim 400$ GHz were obtained
quite early [45], and recently much larger values---for both power and
frequency---have been measured [46]; however, output spectral linewidths
for this configuration are considerably larger than for the resonant
propagation configuration.

The reason for this relative performance difference between the two
configurations is readily understood in an intuitive way: A resonant
propagation oscillator is normally operated in zero, or in any case small,
external magnetic field. Its oscillation frequency is therefore determined
essentially by the junction length and, to a lesser extent, by the dc bias
current. This suggests a rather sharply defined frequency, but, since
radiation is emitted only during a fluxon reflection from the output end
(after a back and forth traversal of the junction), it seems plausible that
the output power and fundamental frequency might be relatively low;
furthermore, for this same reason, the output voltage waveform of a
resonant propagation oscillator consists of a train of fairly sharp spikes,
which suggests that output power is delivered into a broad range of
harmonics rather than being concentrated at the fundamental frequency. In
contrast, the oscillation frequency of a flux flow oscillator depends
strongly on the value of the external magnetic field [21], which means that
there exists another source of fluctuations that might broaden the output
spectral linewidth. However, since fluxons impinge on the output end in a
tightly packed, unidirectional train, as shown in Fig. 2, it seems
plausible that the output power and  fundamental frequency might be
relatively high; furthermore, as is also apparent from Fig. 2, the output
voltage waveform is quite nearly sinusoidal, which suggests that output
power is largely concentrated at the fundamental frequency.

In an effort to boost output power levels and to reduce oscillator
line\-widths, several groups have begun to explore the idea of employing
phase-locked arrays of long-junction oscillators, following the lead of
corresponding work done in the context of small-junction oscillators, in
large measure by the Stony Brook group [47]. So far, more attention has
been dedicated to arrays of resonant propagation oscillators [48-51], in
an attempt to raise output powers to practically useful levels, but quite
recently the problem of phase locking of a flux flow oscillator has also
been addressed [52]. In particular, in the work reported by Monaco {\em et
al.} [50], phase locking of a controlled number, from one to ten, of
series-biased resonant propagation oscillators was obtained at $\sim 10$
GHz by capacitively coupling the oscillator array to a high-Q linear
distributed resonator. The output power---up to $10$ nW in an external,
room temperature, $50$-ohm load---was found to vary quadratically with the
number of locked junctions, as shown in Fig. 5, in accordance with
theoretical expectations.
\begin{figure}
\vspace{8.4cm}
\caption[]{Variation of measured output power into an external $50$-ohm
load from an experimental series-biased, resonant propagation array, with
number of active junctions, $n$. Vertical axis normalized to $P_{1} = 1$.
Points: experimental data; solid curve: parabola $P_{n}/P_{1} = n^{2}$.}
\end{figure}

Monaco {\em et al.} [50] also showed that the behavior of their device can
be described, at least to a first level of approximation, by a simple model
in which the physical, distributed resonator is replaced by a lumped,
linear tank circuit. In this model each junction is coupled to the tank
through a capacitance which also comprises a part of the resonator, and the
dynamics of the oscillating fluxon in each junction is described by the
particle-map approach appropriate to overlap geometry junctions, as
outlined above in Sect. 3. To calculate the effect of the junctions on the
tank one may assume, consistently with the assumed particle nature of the
fluxons, that each fluxon reflecting at the tank boundary can be described
as a delta function voltage pulse of weight $4\pi$. In this way the tank
voltage may be calculated analytically as a sum of impulse responses. The
effect of the tank on each junction is to modify the junction boundary
condition at the tank boundary; specifically, the current fed into each
junction by the tank can be calculated knowing the tank voltage and the
junction coupling capacitance. The resulting model is formally equivalent
to an $(n+2)$-dimensional functional map, in which $n$ is the number of
locked junctions. Although not readily tractable analytically, the system
dynamics can easily be solved by iterating the map numerically. One salient
feature that emerges from such a solution is that the tank voltage
amplitude varies linearly with the number of locked junctions, $n$; this
implies that the output power from the array varies as $n^{2}$, in
agreement with the experimental result shown in Fig. 5.

Although apparently quite promising as an approach to obtain a narrow
linewidth oscillator capable of furnishing a reasonable amount of power,
the use of arrays of resonant propagation oscillators is subject to some
intrinsic limitations regarding the number of junctions that can be locked
together [53]; moreover, the practical problems involved in biasing such an
array in a simple and controllable manner might well limit the feasibility
of this approach. Nonetheless, more work is needed in this direction in
order to fully clarify the situation.

Several attempts to design and construct integrated receivers consisting of
SIS mixers and long-junction oscillators have been described in the
literature. The first of these, to my knowledge, was reported by Crete {\em
et al.} [54] in 1990. This project consisted of a bow-tie antenna, an SIS
mixer, and a $10$-junction, resonant propagation oscillator array, designed
to operate at $94$~GHz, a region of considerable interest for radio
astronomy applications; no experimental measurements on the complete
receiver were reported, however. Zhang {\em et al.} [55] presented results
on a receiver consisting of an SIS mixer and a long-junction oscillator
that could be operated in either the resonant propagation mode or the flux
flow mode, depending on the oscillator junction current density---a
fabrication parameter subject to direct experimental control. Estimated
output powers of $3.2$ nW at $106$ GHz with the oscillator operating in the
resonant propagation mode and $4.6$ nW at $261$ GHz with the oscillator
operating in the flux flow mode were found for this device. In a successive
work Zhang {\em et al.} [56] reported measurements on a very similar
receiver in which two flux flow oscillators were coupled to the same SIS
mixer---one oscillator being coupled fairly strongly and the other fairly
weakly---so as to measure the output linewidths of the oscillators by
mixing them down to a lower intermediate frequency. In this way a composite
linewidth of $\sim 2$ MHz in the band $280-330$ GHz was measured. Moreover,
an available power incident on the mixer of $430$ nW at $320$ GHz was
estimated for this device, which is probably more than adequate for optimal
mixer performance. Also at 1992 ASC, Koshelets {\em et al.} [57] presented
results on two receiver designs, both based on niobium trilayer technology.
The first of these consisted of a flux flow oscillator coupled to a single
small-junction SIS mixer. For this design, an output power of more than
$100$ nW at $256$ GHz was estimated. The oscillator output linewidth was
measured at $82$ and $112$ GHz by mixing the flux flow oscillator signal
with the signal from a backward wave oscillator and analyzing the resulting
intermediate frequency spectrum. A value of $\sim 1$ MHz was found, which,
however, was believed to be an upper limit due to the measurement
instrumentation. The second receiver consisted of a flux flow oscillator
coupled to a multi-junction SIS mixer array. For this design, a minimum
receiver noise temperature of $85$ K at $140$ GHz was measured, with an
oscillator output power of $\sim 20$ nW.

These results, together with those presented by Andy Smith elsewhere in
this volume [58], clearly demonstrate that superconducting technology for
the construction of low-noise, integrated millimeter-wave receivers is
rapidly approaching maturity.
\subsection{Flux Flow Amplifiers}
As mentioned briefly above in Sect. 3.1, that the flux flow mode of
propagation offers potential applications for amplification was recognized
quite early [22]. In the intervening years, much of the pioneering work in
this area has been spearheaded by the group working at the University of
Wisconsin in Madison---though it is certainly true that work by the
Japanese groups has proceeded continuously. In recent times interest in
this application has enjoyed a widespread diffusion, so that today new
results are developing rapidly.

The basic idea underlying the operation of a flux flow amplifier is quite
readily grasped from the general description of the flux flow phenomenon
presented in Sect. 3.1: In the flux flow mode the number of fluxons
injected into the junction per unit time depends essentially on the
strength of the externally applied magnetic field. Since the average
junction voltage is proportional to the product of the number of fluxons in
the junction and their average propagation velocity, an increase of
magnetic field, at constant bias current, causes an increase of the
junction voltage. With an appropriate choice of parameters---in
particular, with the junction biased on a velocity matching step, and with
an appropriate load line---this mechanism can give rise to amplification.

In one of the earliest configurations studied [59], called the `vortex flow
transistor' (VFT), the driving magnetic field was furnished by a
superconducting control line lying either above or below the junction and
electrically insulated from it. In this configuration a time-varying
current in the control line causes a time-varying voltage across the
junction, so that the basic gain mechanism can be characterized as a
transresistance. Properly designed, this device can give significant values
of power gain [60].

An alternative procedure for introducing an input signal into such a device
is to modulate a current injected directly into the junction
counterelectrode [61-63]; this configuration was given the name
`superconducting current injection transistor' (SuperCIT). The basic idea
here is that, as suggested by Eqs. (5) and (6), such an injected current
plays a r\^{o}le in the junction dynamics that is very similar to that
played by a magnetic field. In contrast with the VFT, which is normally
based on a junction having a low-loss, hysteretic current-voltage
characteristic, the SuperCIT is normally based on a junction having a
high-loss, non-hysteretic current-voltage characteristic. For this
configuration, the device transresistance can be enhanced by constructing
the junction with a thin counterelectrode and exploiting the resulting
kinetic inductance [64]. However, the SuperCIT has the disadvantage with
respect to the VFT of being intrinsically a three-terminal, rather than a
four-terminal, device, a fact which leads to obvious grounding problems for
the eventual construction of arrays of such devices.

The performance features of both flux flow oscillators and flux flow
amplifiers can be enhanced if sharp, {\em i.e.}, low dynamic-resistance
velocity-matching steps can be achieved. A number of attempts to realize
such low resistance steps have been based on appropriate tailoring of the
junction geometry, in particular, by the addition of `ears' or `wings' to
the junction structure [21]. An alternative approach, devised by
Nevirkovets {\em et al.} [65], utilizes a dc edge injection current
proportional to the junction bias current, with relative polarities chosen
so that the effect of the injection current is to oppose that of the
applied magnetic field as regards the injection of fluxons into the
junction. The result is a sort of negative feedback which tends to stabilize
the number of fluxons in the junction, and hence the average junction
voltage. Analytical results [65] and numerical simulations [65,66] suggest
that this might prove to be a simple and effective technique for improving
device performance.

A related approach to the realization of flux flow amplification has been
the so-called `superconducting flux flow transistor' (SFFT) [67-69]. This
consists of an array of weak links---not tunnel junctions---patterned into
a single superconducting film, with device input provided by the local
magnetic field generated by a separate control line located near one end of
the array. Although the dynamic behavior of weak link arrays has been
modelled in terms of a discretized version of Eq. (2) [70], it is not clear
that this is entirely correct inasmuch as magnetic flux is carried through
such a discrete array by Abrikosov vortices rather than by Josephson
vortices (see, {\em e.g.}, Likharev [71] for a detailed discussion
regarding the similarities and differences between these two types of
magnetic flux structures).

The original motivation for studying the SFFT was probably the desire to
create novel active electronic applications for the new high-temperature
superconductors, since good tunnel junction characteristics in high-$T_{c}$
materials are as yet difficult to realize (but see the paper by Alex
Braginski [72] for recent progress on this topic); lately, however, the
SFFT has begun to assume an increasing importance in its own right inasmuch
as there are indications that it may display some performance
characteristics that are superior to its long-junction counterparts. An
indicative example is the SFFT amplifier described by Martens {\em et al.}
[73] at 1992 ASC: this device showed a gain of $7$ dB over a bandwidth of
$50$ GHz.
\section{Conclusions}
The central theme of the NATO ASI on which this present volume is based was
that the time now seems ripe for the field of Superconducting Electronics,
which has long played a Cinderella r\^{o}le, to make its debut at the ball.
My task has been to present the notion that devices based on soliton
propagation effects, which have been considered a particularly esoteric
aspect of a generally esoteric field, deserve serious attention in this
context. I have tried to show that the basic ideas involved are, in fact,
quite readily accessible, and that they do offer a rich potential for
practical applications. Here, I have focused on just two fairly simple
analog applications---oscillators and amplifiers---leaving the exciting
area of digital applications to the able exposition of Kostya Likharev
[74]. If I succeed in communicating to the reader even a part of the
enthusiasm that was prevalent during the ASI, then my efforts will have
been amply repaid.
\acknowledgements
The work described herein is largely the fruit of a long-standing and
felicitous collaboration between the Josephson junction groups working at
the University of Salerno, the Cybernetics Institute of the Italian CNR,
and the Technical University of Denmark; I wish to express my gratitude to
my many friends and colleagues working in these groups. Special thanks go
to Valery Koshelets, Jim Nordman, and Yongming Zhang for sending me
pre-publication material regarding their work on flux flow devices, and to
Jim Nordman, Boris Malomed, Niels Pedersen, and Alexey Ustinov for
illuminating comments and a critical reading of the manuscript. Financial
support from MURST and from the Progetto Finalizzato ``Tecnologie
Superconduttive e Criogeniche'' del CNR (Italy) is gratefully acknowledged.
\end{document}